\documentclass{vgtc}                          

\ifpdf
  \pdfoutput=1\relax                   
  \usepackage{graphicx}                
  \usepackage{caption}
  \DeclareGraphicsExtensions{.pdf,.png,.jpg,.jpeg} 
\else
  \usepackage{graphicx}                
  \usepackage{caption}
  \DeclareGraphicsExtensions{.eps}     
\fi%

\graphicspath{{figures/}{pictures/}{images/}{./}} 

\usepackage{microtype}                 
\PassOptionsToPackage{warn}{textcomp}  
\usepackage{textcomp}                  
\usepackage{mathptmx}                  
\usepackage{times}                     
\usepackage{cite}                      
\usepackage{tabu}                      
\usepackage{booktabs}                  
\usepackage{amsmath}
\usepackage{amssymb}
\usepackage[export]{adjustbox}
\onlineid{0}

\vgtccategory{Research}

\vgtcinsertpkg

\title{T2VTree: User-Centered Visual Analytics for Agent-Assisted Thought-to-Video Authoring}

\author{%
Zhuoyun Zheng$^{1,2}$\thanks{%
Affiliations:
$^{1}$ Computer Network Information Center, Chinese Academy of Sciences, Beijing, China;
$^{2}$ University of Chinese Academy of Sciences, Beijing, China;
$^{3}$ Hangzhou Institute for Advanced Study, UCAS, Hangzhou, China;
$^{4}$ University of Technology Sydney, Sydney, New South Wales, Australia.
E-mails:
\{zyzheng, dongyu, chengshiyu, tiandong, sgh\}@cnic.cn;
lianggaorong25@mails.ucas.ac.cn;
liguan@sccas.cn;
\{Jianlong.Zhou, Jie.Liang\}@uts.edu.au. Zhuoyun Zheng and Yu Dong contributed equally to this work. Guihua Shan is the corresponding author.}%
\and
Yu Dong$^{1}$\footnotemark[1]
\and
Gaorong Liang$^{1,2}$
\and
Guan Li$^{1,2}$
\and
Guihua Shan$^{1,2,3}$
\and
Shiyu Cheng$^{1}$
\and
Dong Tian$^{1,2}$
\and
Jianlong Zhou$^{4}$
\and
Jie Liang$^{4}$%
}

\abstract{
Generative models have substantially expanded video generation capabilities, yet practical thought-to-video creation remains a multi-stage, multi-modal, and decision-intensive process.
However, existing tools either hide intermediate decisions behind repeated reruns or expose operator-level workflows that make exploration trace difficult to manage, compare, and reuse.
We present \textit{T2VTree}, a user-centered visual analytics approach for agent-assisted thought-to-video authoring.
\textit{T2VTree} represents authoring process as a tree visualization. Each node in tree binds an editable specification (intent, referenced inputs, workflow choice, prompts, and parameters) with the resulting multimodal outputs, making refinement, branching, and provenance inspection directly operable. To reduce the burden of deciding what to do next, a set of collaborating agents translates step-level intent into an executable plan that remains visible and user-editable before execution. 
We further implement a visual analytics system which integrates branching authoring with in-place preview and stitching for convergent assembly, enabling end-to-end multi-scene creation without leaving the authoring context.
We demonstrate \textit{T2VTreeVA} through two multi-scene case studies and a comparative user study, showing how \textit{T2VTree} visualization and editable agent planning support reliable refinement, localized comparison, and practical reuse in real authoring workflows. T2VTree is available at: \href{https://github.com/tezuka0210/T2VTree}{https://github.com/tezuka0210/T2VTree.}
}


\begin{document}



\firstsection{Introduction}
\label{sec:introduction}
\maketitle

Recent advances in generative video models have substantially expanded video generation capabilities, enabling users to produce short clips from textual descriptions or visual inputs.
While these techniques have demonstrated impressive progress in visual fidelity and motion coherence, producing meaningful video content remains a complex creative process.
Creators rarely begin with a complete and executable specification of the desired video.
Instead, they start from an evolving mental target that must be progressively externalized, structured, and refined through interaction before a satisfactory result can be achieved.

In real workflows, video authoring is inherently multi-stage and multi-modal.
A practical pipeline often begins by establishing visual anchors, such as generating or selecting a reference image, a first frame, or key frames.
Creators then generate motion and scene evolution conditioned on these anchors, and repeatedly adjust prompts, constraints, and parameters to correct undesirable motion, composition drift, or style inconsistency.
Intermediate operations are frequently required, including local edits on images or frames, region-level refinement, and quality enhancement steps such as upscaling or temporal smoothing.
Beyond these operations, meaningful video content commonly requires audio assets, including background music, sound effects, narration, and timing alignment.
Creators finally assemble multiple clips and audio tracks through trimming, stitching, and synchronization to form a coherent result.
This pipeline often spans multiple models and tools, making it essential to coordinate decisions and retain reusable intermediate results across stages.

Importantly, even detailed prompts do not guarantee outputs that match a creator's intent.
Misalignments typically become apparent only after inspecting intermediate results, which forces creators to diagnose deviations and progressively introduce constraints that steer subsequent generations.
As refinements accumulate, creators revisit upstream choices, branch to explore alternatives, compare competing variants, and preserve successful partial decisions for reuse.
Thought-to-video authoring is therefore an exploratory and decision-intensive process.
However, existing systems provide limited support for representing evolving intent at a useful level of abstraction, tracing multi-stage decisions with explicit intermediate results, and managing branching exploration in an inspectable and reusable manner.

Existing tools for generative video authoring commonly follow two usage patterns.
One pattern is end-to-end commercial platforms (e.g., Sora and Kling) that offer streamlined interaction: creators can quickly generate clips from text or images, but intermediate decisions and variants are often difficult to inspect and reuse across iterations.
The other pattern is node-based workflow tools built on open-source model and plugin ecosystems (e.g., ComfyUI), which expose configurable operator graphs and allow creators to integrate specialized operations and community workflows, but require substantial technical expertise and tend to fragment authoring trace across graphs, parameters, and external files. Across both tool types, non-linear exploration remains hard to manage at the level creators actually work.
Creators lack a structured way to keep evolving intent, intermediate states, and their derivation relationships visible and reusable over time.
This lack of process visibility makes it difficult to iteratively refine ideas with confidence, to understand how alternative outcomes relate to prior decisions, and to reliably reuse effective decisions when expanding to additional scenes.

From a visual perspective, thought-to-video authoring can be viewed as a human--AI collaborative exploration problem.
Creators propose and revise candidate directions, treat intermediate results as evidence for decision making, and make iterative choices under uncertainty.
Addressing this problem requires visual representations and interactions that externalize intent, expose decision provenance, and support branch-aware comparison and trace-aware reuse as core authoring operations.

In this paper, we propose \textit{T2VTree}, a user-centered visual analytics approach for thought-to-video authoring.
\textit{T2VTree} externalizes the authoring process as a tree-structured representation in which each node captures a persistent authoring state.
A state co-locates step inputs (e.g., intent, referenced assets, workflow choice, prompts, and key parameters) with the resulting outputs across modalities (image, video, and audio), so that refinement and exploration accumulate without overwriting prior work.
This tree representation makes branching explicit and inspectable, enabling creators to backtrack, compare alternatives under a shared context, and reuse successful partial configurations when expanding to additional scenes.

To make authoring more user-centered and reduce the need for manual workflow assembly, \textit{T2VTree} is coupled with agent-assisted planning that operates at the level of creator-intended authoring actions rather than operator-graph construction.
Creators specify their goals in natural language, optionally providing visual references, and a set of collaborating agents translates these inputs into an actionable next step by selecting an appropriate workflow module and materializing a draft prompt and key parameters for execution.
Instead of treating this automation as a black box, \textit{T2VTree} surfaces the agent-produced plan as node-bound, editable results so creators can review, revise, and reuse decisions in place.

Combined with decision-trace visualization, in-context multi-modal preview, and stitching-based assembly, this design supports human-guided exploration across the full authoring loop, allowing creators to iteratively refine intent, branch and compare alternatives, and converge on a final multi-scene video without leaving the authoring context.

The contributions of this work are as follows:
\begin{itemize}
    \item We report a formative study of experienced creators and distill design requirements for visual analytics support in thought-to-video authoring, including intent externalization, traceable decision provenance, and branch-aware comparison and reuse.
    \item We introduce \textit{T2VTree}, a tree-based visual representation that records persistent states linking step specifications with multi-modal outputs, enabling inspection, branching, localized comparison, and reuse across multi-scene workflows.
    \item We implement \textit{T2VTreeVA}, a visual analytics system that integrates editable agent-assisted planning with branch-aware authoring, in-context multi-modal preview, and stitching-based assembly, and evaluate it through multi-scene usage scenarios and qualitative user study from creators.
\end{itemize}

\section{Related Work}
\label{sec:related_work}
\subsection{Generative multi-modality authoring workflows}
Recent advances in text-to-video and image-to-video generation have driven diverse generative video authoring pipelines in both academia and industry \cite{jan2025text, brooks2024video, blattmann2023stable, team2025kling}.

In practice, creators use visual anchors—such as reference images, first frames, or keyframes—produced by T2I models or derived from existing assets \cite{xiao2025videoauteur, Henschel2025StreamingT2V, liu2025sketchvideo, weng2024art}, and iteratively refine prompts, parameters, or references to mitigate composition drift and style inconsistency \cite{ma2025controllable}. Image generation thus functions both as a prerequisite for video synthesis and as an intermediate representation supporting iterative decision-making \cite{girdhar2023emu}.

Generative video authoring is inherently exploratory and trial-and-error: creators frequently regenerate content, perform localized refinement or quality enhancement on intermediate images or frames \cite{zhang2023adding, mou2024t2i, shi2024dragdiffusion}, and reuse refined results in subsequent stages \cite{Zhang2025LayerCraft}. Temporal coherence is further improved via start/end frame constraints, camera motion specification, or interpolation and smoothing \cite{ma2025controllable}. Creative goals are rarely fully defined at the outset, but clarified through continuous inspection and comparison \cite{cao2025cods}.

Beyond visual content, audio is typically generated independently \cite{peng2025vibevoice, cui2025recent} and synchronized in post-processing \cite{chion2019audio}, with limited support for cross-modal semantic or temporal alignment.

Despite these advances, most generative video tools remain centered on repeated “generate–inspect–retry” cycles \cite{brade2023promptify, trinh2025picture}, treating intermediate results and abandoned alternatives as ephemeral artifacts, which forces creators to rely on memory or external files to manage branching and parameter variations, increasing cognitive overhead \cite{Wang2023Reprompt, suh2023sensecape}.

Node-based or operator-centered authoring interfaces improve controllability by representing generation, editing, and processing steps as composable nodes in executable workflows \cite{callahan2006vistrails, du2023rapsai}, providing expert users fine-grained control \cite{comfyui, nguyen2025genflow, croisdale2025deckflow}. However, they remain focused on execution and computational topology, imposing high cognitive demands \cite{yoghourdjian2020scalability}. Without semantic-level guidance, users must manage complex node semantics, parameter couplings, execution order, and maintain consistency across images, video, and audio. 

Workflow graphs typically encode only dataflow dependencies, so multimodal creative processes are split across subgraphs. Users must infer correspondences between outputs, while evolving intentions, decision rationales, and abandoned alternatives are rarely captured explicitly \cite{dritsas2025multimodal, satkunarajan2025joy, simkute2025ironies}. This fragmentation increases cognitive and management overhead, limiting support for decision-driven, branching exploration. In summary, while existing workflow systems effectively expose execution structures, they remain limited in representing generative video authoring as a creative process centered on decision evolution, branching exploration, and evidence accumulation.

\subsection{Human--AI Collaboration and Creative Visual Analytics}
In the fields of human–AI collaboration and visual analytics, researchers have long investigated how visualization can support user reasoning, exploration, and decision-making in complex and uncertain tasks \cite{eberhard2023effects}. A substantial body of work has shown that creative generation and design activities are inherently non-linear and divergent, characterized by branching exploration rather than single-path optimization \cite{zhang2025auragenome, yu2023user, kolko2018divisiveness, cross2023design, de2025creativity}. However, most conventional tools rely on linear, time-ordered histories, which make early exploration states difficult to revisit, compare, or reuse, thereby constraining parallel exploration and reflective iteration.

Recent visual analytics work has explored how visualizing interaction history supports reflection in generative creation \cite{ragan2015characterizing, croisdale2025deckflow}. PromptThis \cite{guo2024prompthis} visualizes prompt–image relationships to help users review prompt histories, while PromptAid \cite{mishra2025promptaid} supports interactive prompt refinement through coordinated visualizations and perturbation strategies. In broader creative contexts, The Joy of Co-Painting \cite{satkunarajan2025joy} and Sel3DCraft \cite{xiang2025sel3dcraft} demonstrate how provenance- and structure-aware representations can support traceable exploration in image and 3D generation workflows.Together, these studies demonstrate that structured representations of history and branching are critical for reducing trial-and-error costs, supporting iterative exploration, and enhancing users’ sense of control.

Despite these advances, existing work primarily targets data analysis, text generation, or static image and 3D creation, offering limited support for the process management demands of multi-stage, multimodal video generation. Prior research on video visual analytics emphasizes the challenges of reasoning over temporally complex and multimodal video data \cite{he2023videopro, xia2024video}, which often entail higher computational costs, longer latency, and strong cross-stage dependencies. This inherent complexity similarly amplifies the cost of trial and error in generative video workflows \cite{solanki2024script}. In the absence of systematic mechanisms for versioning, branching, and decision tracing, creators are often forced to manually track parameters and intermediate results, increasing both cognitive and operational burden.

This challenge becomes even more pronounced with the introduction of intelligent agents that participate in planning and execution. How to integrate agent-generated suggestions, automated decision processes, and corresponding generative evidence into interpretable and traceable visual structures remains an open problem \cite{heer2019agency, xiang2025promptsculptor, venkatesh2025crea}. Motivated by these gaps, \textit{T2VTree} models generative video authoring as a human–AI collaborative visual analytics process centered on decisions and evidence, rather than solely on execution steps or final outputs.

\section{Formative Study and Design Requirements}
\label{sec:formative}

To ensure that \textit{T2VTree} reflects real authoring practice rather than an idealized one-shot generation pipeline, we conducted a formative study to understand how creators produce short videos with existing generative tools, what recurring steps they perform, and where breakdowns occur. Our goal is not to exhaustively model all production behaviors, but to distill a stable set of common authoring actions that can be supported as first-class interface operations, together with the key interaction gaps that motivate our design.

\subsection{Formative Study Design}
\label{sec:formative_design}

We recruited five participants (P1--P5) with prior experience in AI-assisted content creation: two were experienced with node-based workflows (e.g., ComfyUI), while three primarily used proprietary end-to-end platforms. Each participant was asked to create a short multi-scene video (3--5 scenes) for a realistic topic they might plausibly author (e.g., cultural heritage introduction, travel mood clip, atmospheric micro story). Only a coarse high-level goal (theme and tone) was provided, requiring participants to progressively clarify intent, explore alternatives, and refine intermediate outputs.

Participants used their preferred tools in an open-ended manner, allowing intent to evolve through exploration. Each session lasted approximately 60 minutes, including task execution observation and a semi-structured interview. With consent, we recorded screen activity and collected notes, followed by iterative qualitative coding and author discussion to summarize recurring patterns and breakdowns.

\subsection{Observed Workflow: Common Authoring Actions}
\label{sec:formative_workflow}

Across tools and expertise levels, participants converged on a similar authoring pattern. Rather than generating a complete video in one step, they constructed videos through repeated actions with recognizable inputs and outputs (e.g., textual intent, reference assets, constraints, and generated results).

First, creators established \emph{visual anchors} by generating or selecting reference images to fix composition, style, and atmosphere. Second, they performed \emph{iterative refinement} on images before committing to motion, adjusting prompts and applying intermediate operations such as local edits, structure-preserving refinement, and quality enhancement. Third, creators generated \emph{motion} conditioned on selected anchors and repeatedly corrected temporal issues through re-generation, constraint adjustment, camera control, or frame interpolation. Fourth, creators often produced \emph{audio assets} in parallel and later aligned them with scene timing. Finally, selected clips and audio were \emph{assembled} through trimming, stitching, and synchronization—often requiring export to external tools, increasing overhead and disrupting context.

\subsection{Key Challenges}
\label{sec:formative_challenges}

While the overall action sequence was consistent, participants repeatedly encountered breakdowns in deciding what to do next, managing alternatives, and coordinating tools across modalities. From these sessions, we identify three recurring challenges.

\textbf{C1. Translating intent into the next actionable step is difficult.} Even when participants could articulate detailed prompts, outputs often deviated from intended composition, style, or motion. Existing tools offered limited support for interpreting such deviations and mapping them to concrete corrective actions. As constraints accumulated, creators lacked guidance on whether to revise anchors, introduce controls, or restructure workflows, leading to trial-and-error exploration.

\textbf{C2. Exploration is non-linear, but trace is hard to manage.} Participants frequently generated variants, compared alternatives, and backtracked to earlier results. However, existing systems either hide intermediate steps or expose execution graphs that poorly reflect exploration logic, making variants difficult to relate, compare, and reuse.

\textbf{C3. The pipeline is fragmented across tools and modalities.} Authoring required switching among image refinement, video generation, audio creation, and assembly tools. Repeated asset transfers increased operational burden and caused context loss.

\subsection{Design Requirements}
\label{sec:requirements}

Based on the challenges, we derive four requirements for a user-centered authoring approach that supports thought-to-video creation as iterative, multi-stage exploration.

\textbf{R1. Next-step grounding for intent-to-action translation (addresses C1)}
The system should help creators translate an evolving intent into a concrete \emph{next authoring action} after inspecting intermediate results.
This translation should be grounded in the current context (scene intent, current state, available assets) and produce an actionable draft that creators can revise before execution, reducing ad hoc trial-and-error in selecting corrective operations.

\textbf{R2. Action-level authoring units with clear input--output boundaries. (addresses C1, C3)}
The authoring process should be organized around creator-recognizable actions (e.g., establishing an anchor, refining an image, generating motion with constraints, producing audio, assembling results) with explicit input--output boundaries.
This abstraction allows creators to operate at the level of \emph{what they want to accomplish} rather than operator-graph construction, while remaining stable as underlying workflows or models change.

\textbf{R3. Structure-aware trace for non-linear exploration, comparison, and reuse. (addresses C2)}
The system should externalize the non-linear exploration process as an explicit trace that preserves intermediate states and their derivation relationships.
Creators should be able to backtrack, compare alternatives under a shared context, and reuse successful partial decisions and assets without reconstructing workflows from scratch.

\textbf{R4. Cross-modal continuity across stages and final convergence. (addresses C3)}
The system should maintain continuity across image, video, audio, and assembly stages in a single workspace by preserving consistent state context and supporting explicit asset handoff between actions.
This reduces tool switching and context loss, and supports convergence from exploratory iterations to a coherent multi-scene output.

\label{sec:planning}
\begin{figure}[]
    \centering  
    \includegraphics[width=\linewidth]{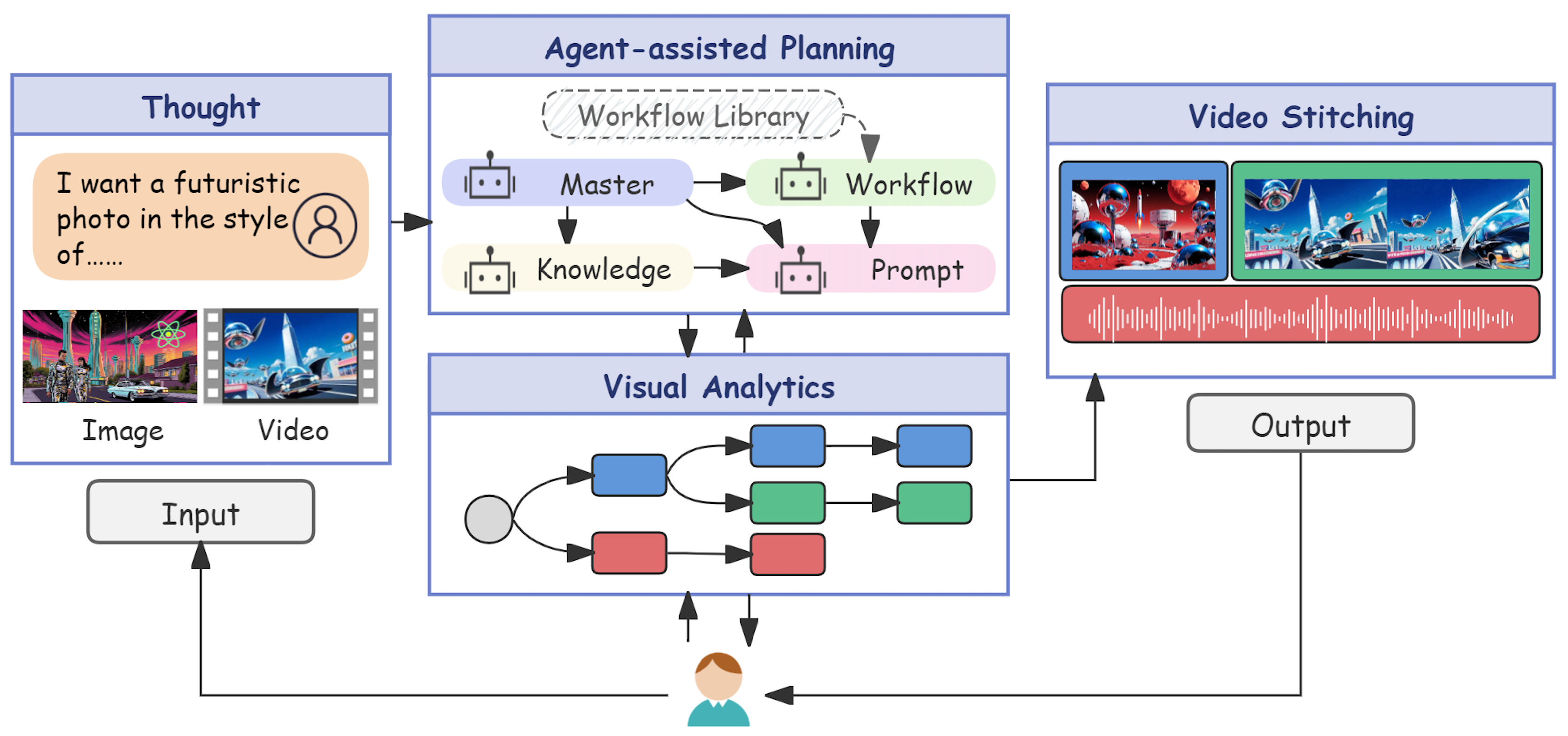} 
    \caption{User-centered authoring loop during video stitiching. Natural-language intent is translated into an editable plan, executed into visual analytics, and curated for video output, enabling branch-and-revise iteration from state to state.}
    \label{fig:agentimage} 
\end{figure}

\section{Authoring Workflow with Agent Assistance}
\label{sec:workflow}

To satisfy the design requirements (R1--R4), we organize thought-to-video authoring as a user-centered cycle (Fig.~\ref{fig:agentimage}) that couples (i) creator intent input, (ii) agent-assisted next-step planning, (iii) visual inspection and revision over persistent states, and (iv) cross-modal convergence through stitching.
Agent assistance primarily addresses next-step grounding (R1) by translating a creator’s intent into an editable action draft, while the interface operationalizes creator-recognizable actions (R2) and preserves an explicit trace for branching, comparison, and reuse (R3).
By keeping image, video, audio, and assembly within a workspace and supporting explicit asset handoff across actions, the cycle also targets cross-modal continuity and final convergence (R4).
The following subsections describe how we define authoring actions, surface planning as editable decisions, and persist authoring states for traceability and reuse.

\subsection{Creator-Recognizable Authoring Actions}
\label{sec:actions}

To support action-level authoring units with clear input--output boundaries (R2), we operationalize recurring creative intents identified in our formative study as a small set of creator-recognizable \emph{authoring actions}. These actions include establishing a visual anchor, refining visual assets, generating motion with constraints, producing audio, and assembling selected assets through trimming, stitching, and synchronization.
Each action is presented through a consistent interface structure and forms a stable unit of progress that remains meaningful to creators even when different models or workflow implementations are used underneath.

Each action is defined by inputs and outputs that align with how creators reason about progress in practice. For example, a creator may provide text with optional reference images to obtain an anchor image, refine an anchor image to improve visual quality, generate a video clip by combining an anchor image (or image pair) with motion constraints, or produce an audio track from a script.
By keeping the interface constant at the action level, this framing supports iterative decision-making: creators can choose to strengthen anchors, revise constraints, or switch action types without being forced to reason about low-level execution structures.

\subsection{Agent-Assisted Planning as Authoring Decisions}
\label{sec:planning}

A central breakdown in practice is deciding the \emph{next actionable step} after inspecting intermediate results.
We address next-step grounding for intent-to-action translation (R1) by introducing agent assistance that produces an \emph{editable plan} before execution.

As illustrated in Fig.~\ref{fig:agentimage}, given a creator's step input consisting of a short intent description and optional visual references, the system constructs context by retrieving the scene-level intent captured at the scene root and the current path context along the active branch.
Four agents then collaborate to translate intent into a concrete next step.
The \emph{Master Agent} consolidates context and proposes an appropriate \emph{action category}.
The \emph{Knowledge Agent} optionally provides brief clarifications for entities or concepts that may affect generation choices.
Given the action category and available inputs, the \emph{Workflow Agent} selects a compatible executable workflow module (identified by a workflow ID).
Finally, the \emph{Prompt Agent} materializes a modality-aware draft specification, including prompt text and key editable parameters aligned with the selected workflow.

Importantly, the plan is not executed as opaque automation.
It is surfaced to the creator as node-bound, inspectable results that can be revised in place, including overriding the suggested action, editing prompts, and adjusting parameters.
Execution is triggered only after review, preserving user control while turning intent-to-action translation into a visible and revisable step.

\subsection{Persisted Authoring State for Traceability and Reuse}
\label{sec:state}

Iterative creation is inherently non-linear: creators branch to explore alternatives, backtrack to earlier stable results, and reuse intermediate assets across steps and scenes.
To support structure-aware trace for comparison and reuse (R3), we treat each step as a persistent \emph{authoring state} rather than a transient execution event.

An authoring state records the creator's intent text, referenced inputs, the selected action and workflow identifier, the agent-produced plan, creator edits to prompts and parameters, and the resulting multimodal outputs (completing one iteration of the cycle in Fig.~\ref{fig:agentimage}).
Each new step is stored as a child of the step it extends, forming an explicit branching structure over states.
This representation supports traceability by preserving decision context, and supports reuse by allowing creators to carry forward effective intermediate assets and configurations without reconstructing workflows from scratch.
Together with in-context preview and stitching-based convergence, persisted states also help maintain cross-modal continuity across stages (R4).

\section{Visual Design of T2VTree}
\label{sec:design}

Guided by the requirements for next-step grounding (R1), action-level authoring (R2), structure-aware trace (R3), and cross-modal continuity (R4), the visual design of \textit{T2VTree} treats thought-to-video authoring as something creators can \emph{see}, \emph{inspect}, and \emph{revise} as they iterate.
Instead of only showing final outputs, the interface makes each step explicit as a persistent state: what the creator requested, what the agents proposed, what was executed, and what was produced.
This enables progressive specification, branch-aware exploration, and reuse without losing context.

\subsection{Design Rationale and Alternatives}
\label{sec:design_alternatives}

\begin{figure}[]
    \centering 
    \includegraphics[width=\linewidth]{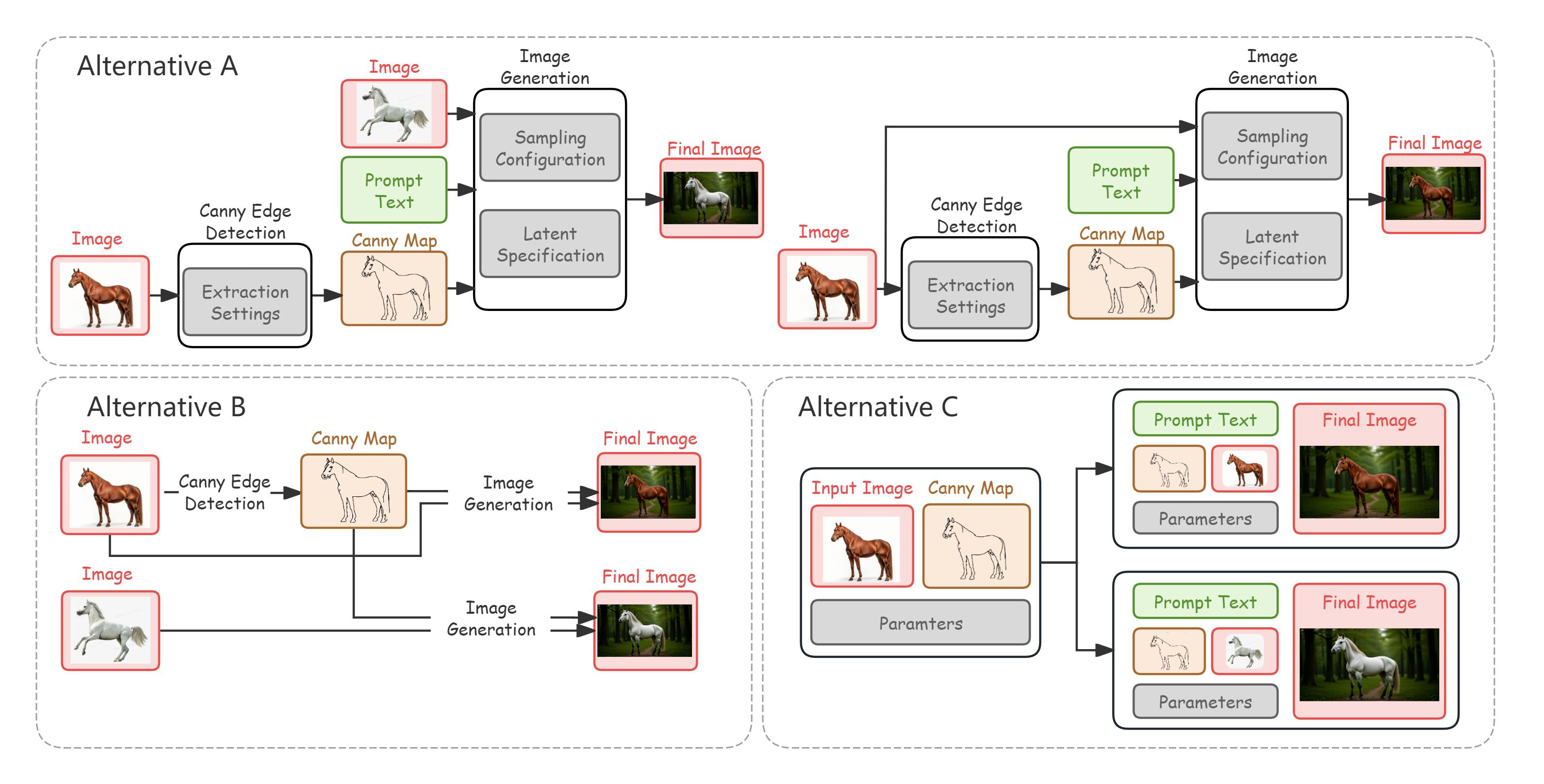} 
    \caption{Three authoring alternatives for structure-guided image generation, motivating the tree-based representation in \textit{T2VTree}.}
    \label{fig:alternative}
\end{figure}

Our representation design evolved through three iterations driven by concrete breakdowns observed during authoring.
We use a structure-guided image generation task as a running example (Fig.~\ref{fig:alternative}): creators derive a structural control signal (e.g., a Canny edge map) from an input image and then generate a constrained result conditioned on both the input and the derived control.

\textbf{Alternative A: Simplifying ComfyUI-style workflow graphs.}
We started from a ComfyUI-inspired representation because node-based workflows are familiar to experienced creators and provide fine-grained control.
To reduce clutter, we grouped low-level operators into higher-level blocks (e.g., \emph{derive structure} and \emph{constrained generation} in Fig.~\ref{fig:alternative} Alternative A).
While this made a single workflow easier to read, authoring progress still relied on graph edits and reruns.
Variants were produced by rewiring blocks or tuning parameters, and the resulting trace was scattered across multiple graph states and exported files.
As a result, comparing alternatives and reusing partial progress required manual bookkeeping, because the graph does not treat a step as a persistent unit of authoring reasoning.

\textbf{Alternative B: Splitting steps into input--output pairs.}
We then attempted a more outcome-oriented view by collapsing each step into an input--output pair (Fig.~\ref{fig:alternative} Alternative B).
This abstraction highlights what inputs led to what results, but it breaks down for control-intensive authoring.
In the example, the control signal is derived from an input image and jointly conditions generation with another image and text.
Once such multi-input dependencies and cross-step reuse are introduced, the structure becomes a DAG with accumulating links, reducing readability and making branching exploration difficult to follow.

\textbf{Alternative C (Final): Action-bounded nodes organized as a tree.}
To keep exploration readable while supporting reuse, we decouple global organization from internal dependencies.
We retain a tree as the global backbone and redefine each node as a self-contained authoring action (Fig.~\ref{fig:alternative} Alternative C).
Each node co-locates the editable specification (intent/prompt, parameters, referenced inputs, and the selected workflow module) with its output, so the node itself becomes the smallest unit for inspection, revision, and comparison.
In this design, child nodes represent subsequent decisions after inspecting a parent outcome, and sibling nodes represent alternatives under the same context.
Reuse is expressed within nodes by referencing prior assets, avoiding explicit dependency edges that would collapse the view into a graph.

This progression preserves creator-recognizable decision boundaries while progressively abstracting away operator-level complexity, leading to the tree representation adopted in \textit{T2VTree}.

\subsection{Node Design}
\label{sec:design_nodes}

\begin{figure}[t]
    \centering
    \includegraphics[width=\linewidth]{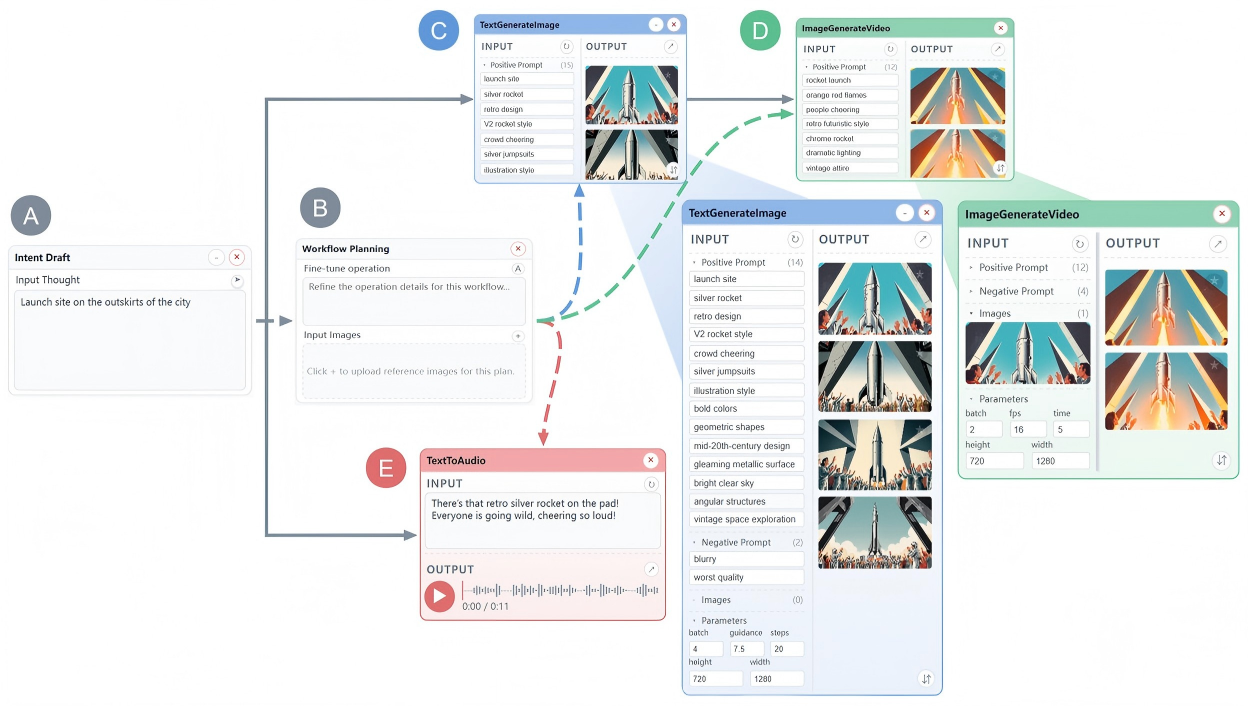}
    \caption{The visual design of \textit{T2VTree}. An intent-first authoring step is materialized as a \emph{Workflow Planning} node that can be transformed into an executable modal color-coded node after agent planning. Solid arrows indicate the user-visible authoring trace after generation, while dashed arrows indicate system-side transformations from planning to modality-specific states.}
    \label{fig:nodeimage}
\end{figure}

We design each node as a \emph{persistent authoring state} that keeps a step understandable \emph{in place}.
As shown in Fig.~\ref{fig:nodeimage}, a state co-locates (i) the user’s intent input, (ii) the editable specification to execute the step (e.g., prompt drafts, parameters, and optional references), and (iii) the resulting multimodal artifacts.
This co-location makes revision and comparison local (R4): creators can adjust what matters for a step without reconstructing context from external sources.

To keep readable as the tree grows, we use a lightweight encoding.
Non-media states (e.g., intent capture and planning) are rendered in neutral tones, while media-producing states are color-coded by output modality (initially image in blue, video in green, and audio in red).
This can help quickly identify \emph{where intent was specified}, \emph{where planning decisions were made}, and \emph{which assets were produced}.

\textbf{Intent Draft.}
The \emph{Intent Draft} node captures scene-level intent as an editable semantic anchor. Creators can input narrative roles, styles, and moods here. Once the intent is finalized, a {\textit{Save} button \includegraphics[height=2.0ex, valign=c]{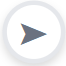} locks this semantic anchor, serving as a stable reference point when creators branch, backtrack, or revise earlier decisions.

\textbf{Workflow Planning as intent-to-action bridge.}
A distinctive design is the \emph{Workflow Planning} node (Fig.~\ref{fig:nodeimage}B), which frames a step as \emph{what the creator intends to achieve next}.Creators provide a short request and may attach visual references (e.g., an image anchor) via an \textit{Upload} button \includegraphics[height=2.0ex, valign=c]{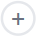}. A \textit{Send} button \includegraphics[height=2.0ex, valign=c]{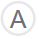} then dispatches the context to the collaborating agents, which materialize a concrete plan, including a suggested authoring action, prompt drafts, and key editable controls.
Crucially, these planning results are \emph{surfaced in the node} so creators can inspect and revise them before committing to execution.

\textbf{Planning-to-modal transformation.}
Rather than remaining as a transient stub, a planning state is \emph{materialized} into a modality-specific node once an action is selected (dashed arrows in Fig.~\ref{fig:nodeimage}).
Depending on the planned operation, the planning node can become an \emph{image} node (blue), a \emph{video} node (green), or an \emph{audio} node (red).
This design keeps the tree trace composed of actionable, comparable states, while still making the intent-to-action translation explicit.

\textbf{Modal nodes with in-place revision.}
Modal nodes share a consistent layout that separates \emph{INPUT} (editable specification) from \emph{OUTPUT} (generated candidates).
For image generation (Fig.~\ref{fig:nodeimage}C), creators can revise prompt drafts, adjust parameters, and optionally provide reference images to refine anchors or perform constraint-guided refinement.
For video generation (Fig.~\ref{fig:nodeimage}D), the node records the chosen image anchor(s) and temporal controls alongside the generated clip, enabling motion-related decisions to remain inspectable and comparable across alternatives.
For audio generation (Fig.~\ref{fig:nodeimage}E), the node couples a narration/script prompt with the produced audio so audio decisions can be revised and reused during multi-scene authoring.
Across modalities, retaining both specification and results within the same node supports iterative tuning while preserving the decision context for later comparison and reuse.
Nodes provide lightweight controls for iteration, including \textit{Generate} \includegraphics[height=2.0ex, valign=c]{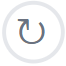}, \textit{vertical resizing} \includegraphics[height=2.0ex, valign=c]{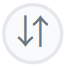} for detailed inspection, and \textit{Add to Stitch} \includegraphics[height=2.0ex, valign=c]{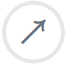} for final composition.

\subsection{Tree Design}
\label{sec:design_tree}

The tree provides a compact global organization for iterative authoring.
Rather than relying on explicit edge meanings, the structure is read through its hierarchy: depth reflects progressive refinement within a creative direction, while siblings capture parallel attempts and alternative options under a shared context.
All branches originate from a single \emph{Init} root that represents the project start state.

\textbf{Parent--child relation: progressive refinement.}
A parent--child relation represents stepwise deepening of the current creative direction.
As creators inspect intermediate results, they introduce additional constraints, change the action type, or extend existing outputs into the next stage.
This produces a deeper node that remains grounded in the earlier context, enabling the workflow to be understood as a progressive specification process.
In practice, depth aligns with a common trajectory of authoring: starting from coarse intent, establishing visual anchors, refining them, and then extending them into time-based or compositional outcomes.

\textbf{Sibling relation: parallel paths and alternatives.}
Siblings represent concurrent directions explored from the same point in the process.
They externalize alternatives as first-class states, so creators can compare competing variants without losing earlier results (R3).
Siblings are not restricted to one modality or one action type: parallel attempts may differ in prompts, references, parameters, or even in the chosen authoring action, yet remain comparable because each node co-locates what was specified and what was produced.

\textbf{Project-root branching for multi-scene creation.}
Multi-scene authoring requires working on several scene directions in parallel.
We therefore organize the workspace around a single \emph{Init} root, from which creators can branch out multiple scene directions as independent top-level paths.
Each scene begins with an \emph{Intent Draft} node that serves as the scene anchor, and subsequent refinements, generation steps, and asset derivations deepen along that scene's branch.
This organization avoids interleaving unrelated iterations into a single linear trace, while keeping all scene directions visible within one workspace for cross-scene navigation.

\textbf{Managing tree with collapsing and pruning.}
As exploration proceeds, local neighborhoods can grow rapidly.
To keep the \textit{T2VTree} readable, branches can be collapsed by \textit{collapse} button \includegraphics[height=2.0ex, valign=c]{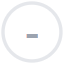} to summarize completed directions while preserving access to internal states when needed.
Unwanted attempts can be removed by \textit{delete} button \includegraphics[height=2.0ex, valign=c]{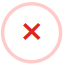} (and their descendants), allowing creators to prune low-value branches without affecting other directions.
Together with an automatic layout adjustment that reclaims screen space, these operations keep the workspace navigable while maintaining a faithful record of refinement and exploration.

\begin{figure*}[]
    \centering
      \includegraphics[width=\linewidth]{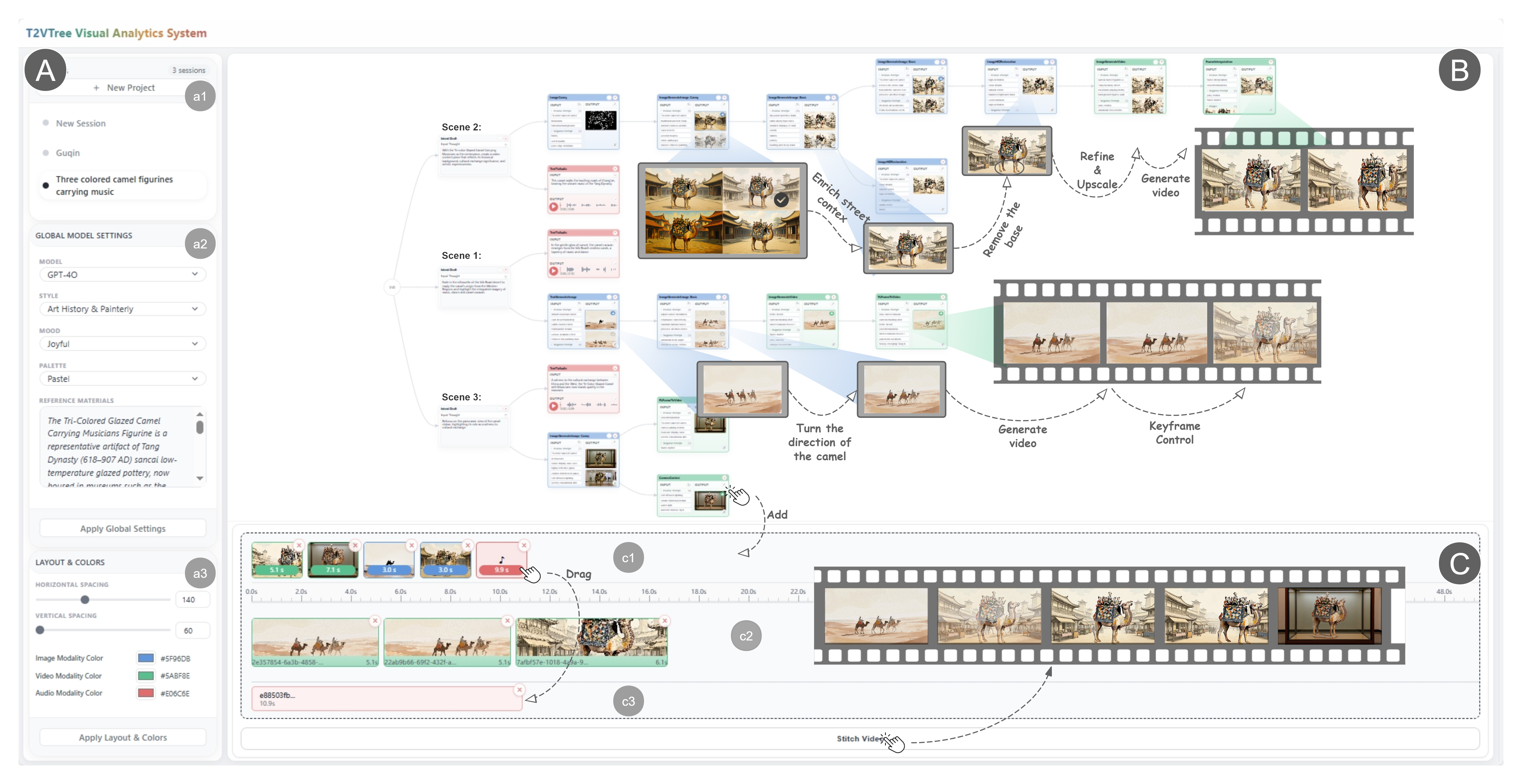}
      \caption{The \textit{T2VTreeVA} interface showcasing the authoring session for Case 1. (A) Control Panel used to input the global historical context and style settings. (B) T2VTree View visualizing the branching provenance trace, where nodes represent generated image/video/audio and sibling branches show alternative processes and refinements. (C) Video Stitching View assembling selected clips from divergent branches into the final documentary timeline.}
      \label{fig:T2VTree}
\end{figure*}

\section{T2VTree Visual Analytics System}
\label{sec:system}

The \textit{T2VTree} visual analytics system operationalizes the authoring abstractions in Sections~\ref{sec:workflow} and~\ref{sec:design} as an interactive workspace for thought-to-video authoring. It maps authoring intents and persisted authoring states into coordinated views that support inspection, revision, branching exploration, and stitching output.

\subsection{Control Panel}
The Control Panel (Fig.~\ref{fig:T2VTree}A) provides global settings shared across nodes and scenes, covering project scope, agent context defaults, and tree layout/encoding.

\textbf{Project management.}
The top (a1) supports creating, deleting, and switching projects; each project maintains an independent tree.

\textbf{Global context defaults.}
The middle (a2) configures the base agent model and session-level context (style, mood, palette), with optional background text or reference materials.
These settings are automatically injected into agent-assisted planning for newly created nodes, while remaining locally overridable by node-level intents, prompts, or parameters.

\textbf{Layout and modality encoding.}
The bottom (a3) adjusts global node spacing (horizontal/vertical) and exposes the modality color mapping (image/video/audio) for customization, ensuring consistent encoding in the T2VTree View.

\subsection{T2VTree View}
The T2VTree View (Fig.~\ref{fig:T2VTree}B) is the primary workspace for inspecting, branching, and revising authoring states. It renders the tree on an infinite canvas to support large-scale exploration.

\textbf{Node cards as persisted authoring states.}
Each card corresponds to a persisted authoring state, co-locating step-level intent, editable inputs (prompts, parameters, references), and the resulting multimodal outputs. This enables in-context inspection and revision without reconstructing state from external tools.

\textbf{Direct manipulation and branching.}
Selecting a node sets it as the source state and spawns an \emph{Add Workflow} placeholder as its child for specifying next-step intent.
After agent-assisted planning, the placeholder is replaced by a concrete media node that reflects the selected authoring action.
This interaction makes branching semantics explicit: child nodes encode refinements/extensions, while sibling nodes encode alternative decisions under the same parent context.
Nodes can be collapsed or expanded to manage visual complexity, preserving overview while retaining access to detailed histories when needed.

\textbf{Integrated multi-modal preview.}
Outputs are previewed within node cards: images and videos are shown as thumbnails, videos loop for rapid scanning, and double-click opens higher-resolution previews.
Audio nodes visualize waveforms and support in-place playback, enabling quick assessment without leaving the tree.

\subsection{Video Stitching View}
The Video Stitching View (Fig.~\ref{fig:T2VTree}C) supports the transition from divergent exploration to convergent assembly through a multi-track timeline editor.

\textbf{Candidate collection.}
Creators can collect images, video clips, and audio tracks from any branch into a temporary collection area (c1). Candidates retain modality color encodings to maintain consistency with the tree.

\textbf{Drag-and-drop for stitching.}
Collected assets are dragged onto timeline tracks (c2, c3) and reordered to form a linear narrative. The timeline provides basic alignment and ordering controls for synchronizing visuals and audio.

\textbf{End-to-end traceability.}
Assembled segments remain linked to their originating authoring states, enabling creators to jump back to the corresponding nodes, revise prompts/parameters, regenerate assets, and update the stitched result while preserving provenance.

\section{Evaluation}
\label{sec:evaluation}
We report two case studies and user study to demonstrate how \textit{T2VTreeVA} supports end-to-end thought-to-video authoring in realistic multi-stage settings.

\subsection{Case 1: From Thoughts to Structured Video Authoring}
\label{sec:case1}

We constructed a short video about the \textit{Tricolor Glazed Pottery Camel} to demonstrate how \textit{T2VTreeVA} supports multi-scene, multi-modal authoring from a vague idea to a stitched final video.
Figure~\ref{fig:T2VTree} shows the authoring trace in Case~1 as a tree grown from an \textit{Init} root, where each node encapsulates its step-level specification and the generated outputs.

\textbf{Scene-level setup and intent drafting.}
We first created a new project and configured global generation preferences (Fig.~\ref{fig:T2VTree}A), including the model and a coarse style/mood preset (a2), as well as the modality color mapping (a3).
We also entered the background information, including its dynasty and symbolic significance, into the global \textit{Reference Material} area, establishing the thematic keynote of cultural exchange between Chinese and Western civilizations in the Tang Dynasty. We then authored three scenes by expanding three branches from the \textit{Init} node (Fig.~\ref{fig:T2VTree}B, ``Scene 1--3'').
For each scene, we began with an intent card that describes the desired content and constraints (e.g., historical tone and scene theme), serving as the editable entry point before any generation.

\textbf{Agent-assisted workflow planning and prompt materialization.}
For the main scene depicting the camel in Tang Dynasty Chang'an, we invoked agent assistance on the planning node.
The system returned (i) an editable prompt draft and (ii) a selected workflow module that matches the intent (e.g., structure-guided image generation).
These planning outputs are stored in the node state and rendered as a workflow-styled card, allowing us to modify prompts/parameters before execution.
This step eliminates the need to manually assemble low-level operator graphs while keeping the selected workflow and prompt explicit in the trace.

\textbf{Iterative image refinement as traceable branches.}
Starting from an initial anchor image, we first examined four generated stylistic variants and selected the light watercolor sketch style as the established visual baseline. As illustrated in Fig.~\ref{fig:T2VTree}B, alternative refinement attempts are recorded as sibling image nodes under a shared parent node. The parent node preserves the content that remains unchanged during refinement, while each sibling node captures a distinct revision outcome.

In this scene, we created one branch that directly applies an upscaling operation (``Refine \& Upscale'') to quickly assess the visual quality of the selected style. In parallel, another branch performs two localized refinements commonly used in heritage-style authoring: enriching the scene context by adding background details (``Enrich street contex'') and removing redundant elements ``Remove the base''). We then applied an upscaling step as a derived child node (``Refine \& Upscale'') to improve the quality of the selected anchor image before proceeding to motion generation.

\textbf{Image-to-video handoff and video-side correction.}
Once the final anchor was selected, we generated a video clip, represented as a green node derived from the blue image node (Fig.~\ref{fig:T2VTree}B, ``Generate video'').
When the first clip exhibited temporal results, we created a derived video node applying frame interpolation while retaining the original clip in the parent state for fallback.

\textbf{Multi-scene construction and transition.} To expand the narrative, we initiated a parallel branch for a second scene: a camel carrying musicians in the desert, maintaining the established watercolor style. Observing the movement direction contradicted the narrative (West-to-East), we created a corrective sibling node to adjust the orientation. Deciding to position this as the opening scene, we utilized start-end frame conditioning—using the camel clip's last frame and the street scene's first frame—to generate a transition video node that smoothly bridges the two distinct branches.

\textbf{Audio generation and end-to-end stitching.}
In parallel, we generated an audio narration track as a red node associated with the same project context.
To compose the final documentary, we added selected video clips (from different scene branches) and the chosen audio track into the stitching panel (Fig.~\ref{fig:T2VTree}C).
The stitching view provides a multi-track timeline buffer where creators drag selected clips into tracks (c1--c3) and export a single final video, while each segment remains traceable to its originating node in the tree.

\subsection{Case 2: Multi-Scene Authoring through Progressive Refinement}
\label{sec:case2}
\begin{figure}[]
    \centering 
    \includegraphics[width=\linewidth]{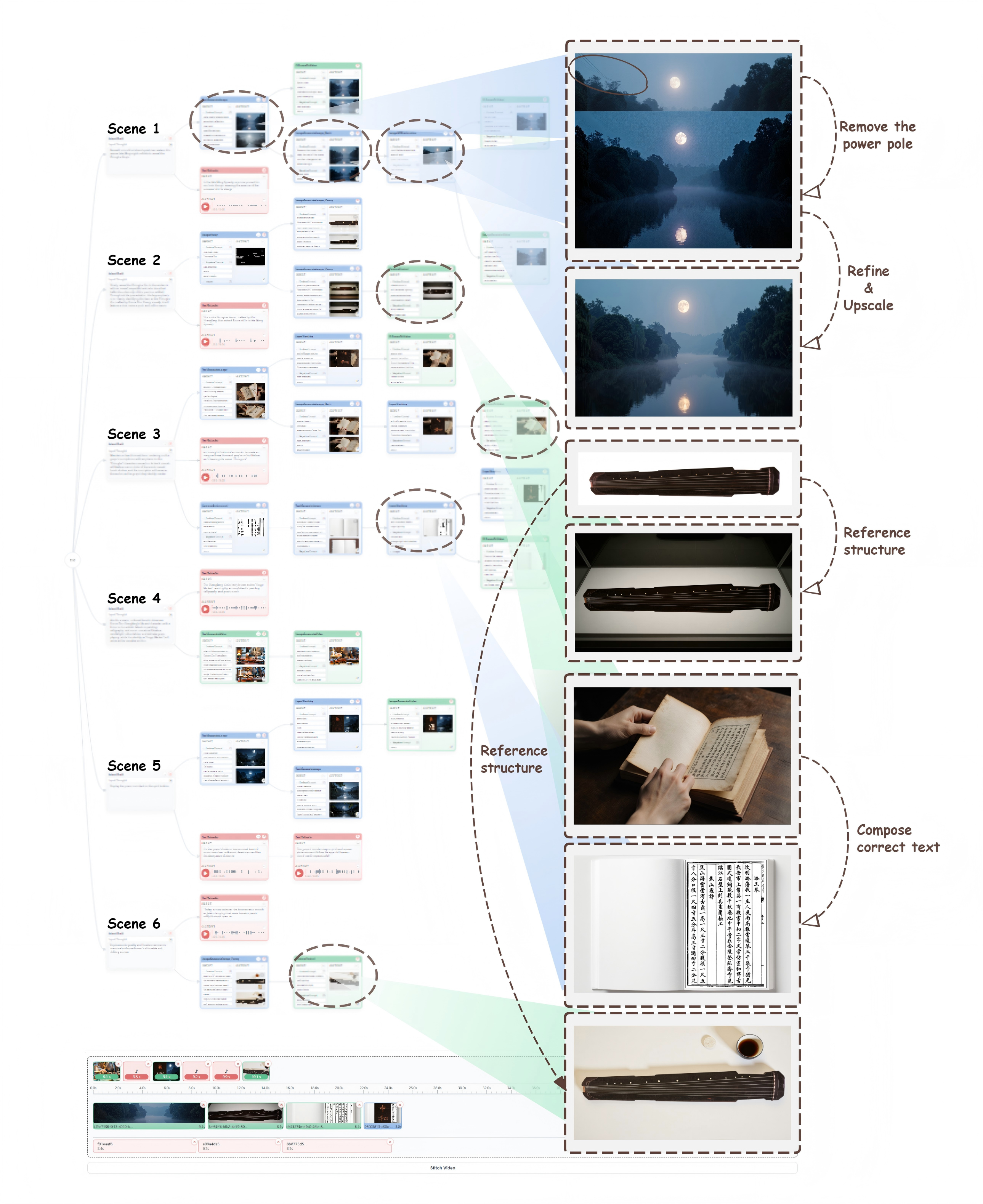} 
    \caption{Case~2 authoring trace visualized in \textit{T2VTree}. Left: The complete multi-scene tree rooted in Init. Right: Magnified views of four key episodes (linked by dashed lines) demonstrating the system's support for progressive refinement.}
    \label{fig:case2}
\end{figure}

The second case emphasizes progressive refinement across multiple scenes authored in parallel, where creators repeatedly correct, tighten, and reuse decisions while keeping the long-range narrative coherent.
We produced a short cultural-heritage video about the Zhonghe Guqin (a Ming-Dynasty instrument associated with Prince Zhu Changfang).
Fig.~\ref{fig:case2} visualizes the process as multiple scene branches grown from the Init root: within each scene, child nodes capture stepwise tightening of decisions, and sibling nodes capture parallel attempts under a shared parent for localized comparison.

\textbf{Progressive Refinement without Restarting.}
We began with reference materials and an initial global preference of ``realistic and photographic.''
For the opening scene, we supplied a rough creative intent (``a quiet atmosphere with trees, river, and moon''), upon which the agent-assisted planning suggested several prompt phrases. After confirming these suggestions, the system generated multiple candidate images within a single node for our selection.
After selecting a candidate with the desired tone, we observed an anachronistic object (a utility pole) that conflicted with the historical setting.
Rather than re-running the scene from scratch, we issued a natural-language correction request to remove the object while preserving composition.
This revision was recorded as a child node, and parallel attempts were expressed as sibling nodes under the same parent so that the correction remained localized and auditable.
We then applied upscaling as a further child node when the edited output became overly soft, while keeping the pre-upscale result available as its parent state.

\textbf{Refining Exhibition Attributes with a Fixed Structural Anchor.}
For the museum scene, we introduced a reference photo of the Guqin and described the goal as keeping the instrument faithful while placing it in an exhibition setting.
As in Case~\ref{sec:case1}, we used ControlNet-Canny guidance to keep geometry stable across iterations (``Reference structure''), so subsequent refinements primarily affected exhibition attributes rather than instrument structure.

Based on fixed structural anchors, we created multiple parallel branches at the same level to explore underspecified attributes. In contrast to strongly constrained branches (e.g., white walls, white lighting, white display stand), we introduced open-ended generation branches with minimal manual intervention, encouraging diversity and avoiding premature commitment to a single direction.
After choosing a preferred direction (e.g., a white stand against a dim background), we tightened the request by explicitly adding a ``glass display cabinet'' constraint, recorded as a child node to keep the decision change explicit in the trace.
We then generated a short close-up clip via a camera zoom workflow, recorded as a green video node derived from the chosen blue image node, making the image-to-video handoff straightforward to verify and revisit.

\textbf{Layered Composition with Traceable Inputs.}
A third scene required presenting historical documentation, where directly generated text was not reliably legible.
We therefore adopted a layered composition strategy that separates content preparation from visual framing.
We first cleaned a screenshot of real historical text (background removal and watermark cleanup), then generated an open-book carrier image, and finally overlaid the cleaned text and the ``Zhonghe'' inscription onto the book (``Compose correct text'').
Each stage was captured as a child node so the composite result remained traceable to concrete inputs and could be revised by replacing a single step rather than restarting the scene.
To connect scenes smoothly, we generated a transition clip with start--end frame control, anchored on selected upstream results rather than overwriting earlier attempts.

\textbf{Backtracking and Reuse as Local Decisions.}
A non-linear episode occurred when depicting the poem engraved on the underside of the Guqin.
We first requested a restrained background (``moonlight, river water, and dew''), then tightened the request to emphasize ``dew drops.''
While the refined result became richer, it also became visually cluttered.
Using the visible decision trace in Fig.~\ref{fig:case2}, we backtracked to an earlier parent state and pursued an alternative direction by overlaying the poem text onto the earlier, cleaner background before converting it into video.
The competing directions remained as sibling nodes under a shared parent, enabling comparison without discarding either attempt.

We can reuse partial progress.
For example, we selected a previously derived structural control as a referenced input inside a new node to request a modern minimalist re-rendering, and then applied a counterclockwise camera rotation workflow to conclude the narrative.
Because reuse is expressed within nodes (as referenced inputs) rather than by adding new edges, the overall trace in Fig.~\ref{fig:case2} remains readable as it grows.

\textbf{Final Video Output.}
Finally, we curated the selected clips (and accompanying audio where applicable) from multiple scene branches into the stitching view for composition.
Because each selected segment remained attached to its originating node, we could verify provenance during assembly and revise upstream decisions when needed without reconstructing intermediate steps.

\subsection{User Study}
\label{sec:userstudy}

We conducted a small-scale comparative study to examine whether \textit{T2VTreeVA} reduces coordination overhead and better supports branch-aware exploration and reuse in thought-to-video authoring, compared with a node-based workflow baseline (ComfyUI).

\textbf{Participants and Design.}
We recruited six participants (P6--P11) with prior experience using ComfyUI for diffusion-based image and video generation; none had used \textit{T2VTreeVA} before.
Participants were evenly assigned to two between-subject conditions: ComfyUI (control) and \textit{T2VTreeVA} (experimental).

\textbf{Task and Procedure.}
Participants authored a five-scene short video under the open-ended theme \emph{``Future City''}.
They were free to iterate, branch, and reuse intermediate results while maintaining a coherent multi-scene outcome.
To capture early-stage authoring behavior, we additionally recorded the elapsed time when participants completed the first three scenes.
Each session lasted approximately 90 minutes, including a 20-minute guided tutorial and a brief warm-up (not measured), followed by task execution and a post-task interview.
We measured task time from the start of scene authoring to successful final export.
During the task, participants were asked to describe their intent and decisions while working; screen interactions and audio comments were recorded.

\textbf{Baseline and Environment.}
To avoid unrealistic baseline, ComfyUI participants were provided with a curated library of workflow templates covering the same major authoring operations supported in \textit{T2VTreeVA}.
Participants did not need to construct operator graphs from scratch, but were responsible for selecting templates, wiring inputs, tuning parameters, and adapting workflows as constraints evolved.
Both conditions shared the same backend server with NVIDIA H100 GPU (80 GB) for media generation, and were operated on a workstation with Intel Xeon Silver 4110 CPUs (2.10 GHz), 64 GB RAM, and NVIDIA Quadro P5000 GPU (16 GB), ensuring comparable execution latency across conditions.
Final assembly differed between conditions: \textit{T2VTreeVA} supports in-system stitching and export, whereas ComfyUI participants assembled and exported clips using their usual workflow outside ComfyUI.

\textbf{Measures.}
We recorded time to complete three scenes ($T_3$) and time to final export ($T_5$).
To attribute time differences beyond model latency, we logged total system waiting time $T_{\text{wait}}$ (accumulated from issuing a request to previewable results) and computed active time $T_{\text{active}} = T_5 - T_{\text{wait}}$ as a proxy for interaction and coordination effort.
To isolate final assembly, we recorded assembly/export time $T_{\text{assemble}}$ (from entering the final selection/assembly stage to export) and computed iterative authoring time $T_{\text{author}} = T_5 - T_{\text{assemble}}$.
We also logged the number of generation calls $N_{\text{calls}}$ (including post-processing) and the number of retained variants $N_{\text{variants}}$ (candidates explicitly kept for comparison or reuse).

Participants additionally completed three Likert-5 questions (1=strongly disagree, 5=strongly agree) about the tool they used:

\textbf{L1: Next-step clarity.} ``This tool helped me turn an initial idea into a clear next step to proceed.''

\textbf{L2: Iterative control.} ``I could iteratively improve my authoring items while keeping the process under control.''

\textbf{L3: End-to-end convergence.} ``I could efficiently converge on and export a final multi-scene video.''

\begin{figure}[t]
\centering

\begin{minipage}{\linewidth}
\small
\setlength{\tabcolsep}{6pt}
\captionof{table}{Quantitative comparison of authoring efficiency and exploration behavior between ComfyUI and \textit{T2VTreeVA}.}
\label{tab:userstudy_metrics}
\begin{tabular}{lcc}
\toprule
\textbf{Metric (Average)} & \textbf{ComfyUI} & \textbf{T2VTreeVA} \\
\midrule
\multicolumn{3}{l}{\textbf{Completion time}} \\
Time to 3 scenes $T_3$ (min) & 34.5 & \textbf{31.3} \\
Time to final export $T_5$ (min) & 68.2 & \textbf{52.6} \\
Assembly and export time $T_{\text{assemble}}$ (min) & 10.5 & \textbf{1.8} \\
\midrule
\multicolumn{3}{l}{\textbf{System waiting vs.\ User effort}} \\
Total waiting time $T_{\text{wait}}$ (min) & 30.4 & \textbf{27.5} \\
Active time $T_{\text{active}}$ (min) & 37.8 & \textbf{25.1} \\
\midrule
\multicolumn{3}{l}{\textbf{Exploration behavior}} \\
Generation calls (counts) $N_{\text{calls}}$ & 32.3 & \textbf{20.3} \\
Retained variants (counts) $N_{\text{variants}}$ & 4.3 & \textbf{8.3} \\
\bottomrule
\end{tabular}
\end{minipage}

\vspace{6pt}

\begin{minipage}{\linewidth}
\centering
\includegraphics[width=\linewidth]{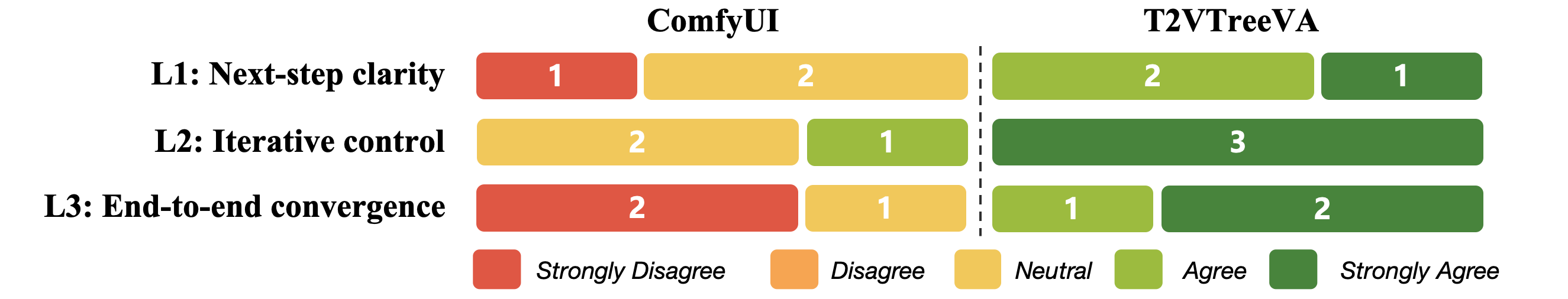}
\caption{The Likert-5 comparative results for L1--L3 between ComfyUI and \textit{T2VTreeVA}.}
\label{fig:likert}
\end{minipage}

\end{figure}

\textbf{Results.}
Across results, total waiting time $T_{\text{wait}}$ was comparable between conditions, while differences in end-to-end completion time ($T_5$) were primarily reflected in $T_{\text{active}}$ and $N_{\text{calls}}$, suggesting reduced interaction and rerun overhead with \textit{T2VTreeVA}.
ComfyUI sessions exhibited higher $N_{\text{calls}}$ in our study, consistent with repeated reruns during parameter tuning and workflow adaptation, whereas \textit{T2VTreeVA} participants more often preserved intermediate states and retained explicit candidates for branch-aware inspection, reflected in higher $N_{\text{variants}}$.
We also observed that $T_{\text{assemble}}$ contributed non-trivially to end-to-end time in the ComfyUI condition, consistent with the need to perform final selection and export outside the authoring workspace.
Likert responses further contextualize quantitative trends (Fig.~\ref{fig:likert}).
For L1, ComfyUI ratings were concentrated around neutral or negative responses, whereas \textit{T2VTreeVA} responses shifted toward agreement, indicating clearer guidance from an initial idea to actionable steps.
For L2, responses under \textit{T2VTreeVA} were consistently positive, while ComfyUI responses were more mixed, reflecting greater effort required to manage iterative refinements.
For L3, ComfyUI received predominantly negative ratings, whereas \textit{T2VTreeVA} responses clustered toward agreement, aligning with the reduced assembly time $T_{\text{assemble}}$ and lower active effort $T_{\text{active}}$ observed in Table~\ref{tab:userstudy_metrics}.
Taken together, these subjective assessments align with the measured differences in coordination overhead and exploration behavior, suggesting that explicit decision traces and in-system assembly support can improve creators' perceived control and convergence in thought-to-video authoring workflows.

\section{Discussion and Limitations}
\label{sec:discussion}

\subsection{Sustainability Considerations}

Sustainability in generative authoring systems depends not only on computational and economic cost, but also on responsible use and reproducible operation under evolving model ecosystems.

\textbf{Cost.}
\textit{T2VTree} uses commercial large language model APIs only for low-frequency planning and prompt materialization, rather than placing them inside high-frequency generation loops.
In our current implementation, a typical planning request consumes 2k--4k tokens, which is small relative to the cost of video generation.
All image, video, and audio generation is executed with open-source models.
This reduces recurring API dependence and improves reproducibility and scalability.
It also supports long-term evolution: as generators improve, \textit{T2VTree}'s state-based representation and workflow-level planning remain applicable, allowing the system to incorporate updated open-source models without changing the core authoring mechanism.

\textbf{Ethics.}
To mitigate ethical risks in generative authoring, \textit{T2VTree} adopts role-based agent assignments and curated prompt templates that constrain downstream generation requests.
These templates encode explicit content rules and context-specific restrictions, providing a lightweight safeguard layer before generation.
At the same time, planning outputs remain visible and editable, and creators must review each step, supporting accountable use without relying on opaque automation.

\subsection{Limitations and Future Work}

\textbf{Limited fine-grained control and iteration overhead.}
In practice, creators may need multiple iterations when generation controls are not sufficient to express a desired constraint or to correct subtle mismatches in composition, motion, or style.
\textit{T2VTreeVA} addresses this by preserving intermediate states and enabling branch-aware comparison, so creators can explore alternatives without losing context.
However, this design can still translate unmet control needs into additional branching and node growth, increasing interaction overhead and reducing authoring fluency.
Future work will explore richer control representations and interaction support that make controllable factors more explicit at each step and reduce unnecessary iterations, such as structured prompt editing, constraint-focused controls, and lightweight guidance based on prior successful states.

\textbf{Scalability for long and complex authoring.}
The tree representation is effective for small- to medium-scale exploration, but long videos and complex storyboards can produce deep and wide trees that are harder to read and navigate.
While \textit{T2VTreeVA} supports collapsing subtrees, future work is needed to improve scalability through higher-level summarization and organization, such as grouping similar variants, highlighting salient paths, and providing multi-level overviews that preserve access to detailed node evidence when needed.

\section{Conclusion}
\label{sec:conclusion}

In this paper, we reframe thought-to-video authoring as a user-centered visual analytics problem in which creators iteratively translate evolving intent into concrete actions, branch to explore alternatives, and converge across modalities.
To support this process, we present \textit{T2VTree} that organizes authoring around action-level units with clear input--output boundaries and externalizes each step as a persistent authoring state.
States co-locate intent, referenced assets, workflow choice, prompts, and key parameters with generated multimodal results, enabling in-place evidence while preserving tree structure for traceability, comparison, and reuse.

\textit{T2VTreeVA} further couples the multiple views with agent-assisted planning that produces an editable plan grounded in the current scene intent and branch context.
By surfacing proposed actions and prompt/parameter drafts as node-bound artifacts for review and revision before execution, the system turns intent-to-action translation into a controllable authoring decision, and then materializes it into modality-specific states that remain inspectable over time.
Together with an integrated stitching workspace, creators can iteratively refine assets across scenes and assemble a coherent video without leaving the authoring context.

We evaluated \textit{T2VTreeVA} through two multi-scene cultural-heritage case studies and a comparative user study.
Across these evaluations, creators were able to progressively refine ideas into complete multi-scene videos, using the tree trace to backtrack, compare alternatives, reuse intermediate results, and coordinate cross-modal handoffs; collectively, the visual analytics design and agent-assisted planning reduced coordination overhead in practical authoring workflows.


\acknowledgments{
This work was supported by Chinese Academy of Sciences–Hunan Province Joint Research Program (Key Technologies Research and Flagship Application for Digital Cultural Heritage, No. 2024JK4002).}

\bibliographystyle{abbrv-doi}

\bibliography{template}

\end{document}